# Superconductivity in Sulfur-Doped Amorphous Carbon Films

*I. Felner, O. Wolf, O. Millo*

*Racah Institute of Physics and the Center for Nanoscience and Nanotechnology, The Hebrew University of Jerusalem, Jerusalem, 91904, Israel*

**Abstract**

Following our previous investigations on superconductivity in amorphous carbon (aC) based systems; we have prepared thin composite aC-W films using electron-beam induced deposition. The films did not show any sign for superconductivity above 5 K. However, local, non-percolative, superconductivity emerged at $T_c$ = 34.4 K after treatment with sulfur at 250 ºC for 24 hours. The superconducting features in the magnetization curves were by far sharper compared to our previous results, and the shielding fraction increased by about an order of magnitude. Our data suggest that pairing and localized superconductivity take place in the aC-S regions, whereas phase coherence, assisted by the W inclusions, was enhanced compared to our previous samples, yet still not to the degree of achieving global phase-coherence and percolating superconductivity.

## 1. Introduction

The possibility of having high temperature superconductivity (SC) in graphite and at its interfaces [1,2], as well as in disordered amorphous carbon [3] has been the subject of only a few experimental studies over the last 5 years. Superconductivity was observed either by doping the graphene layers and/or at the graphite surface region due to their topologically protected flat bands [4-8]. The recent report of granular room temperature SC in water-treated graphite powder [9] has encouraged more intense activity in this exciting area.

Amorphous carbon (aC) materials with dimensions intermediate between molecules and bulk solids are intriguing, due to their novel electronic features which depend on various parameters such as the particle size, topology and surface conditions. Typical aC powders and/or aC thin films have a mixture of $sp^2$ and $sp^3$ bonding. Changing the $sp^2/sp^3$ ratio tunes the band-gap of aC between that of diamond (100% $sp^3$, band-gap $E_g = 5.4$ eV) and graphite (100% $sp^2$, semimetal with $E_g=0$) [10]. Amorphous carbon contains partially graphitized carbon fragments that possess both negative and positive curvatures; therefore, aC is a promising candidate for searching for SC at elevated temperatures.

In recent studies we have shown, for three different systems, that tiny fractions of localized SC phases can be observed in aC treated with sulfur. (i) We have demonstrated that samples extracted from an old *commercial* inhomogeneous aC powder exhibit traces of non-percolating SC onset at $T_c$ ~32 K and $T_c$ ~ 65 K [3,11]. Interestingly, the SC shielding fraction (SF) was enhanced when this aC powder was annealed with sulfur at 250 ºC [3]. (ii) These results encouraged us to perform comprehensive investigations, in a *synthetic* aC powder obtained by melting of pure sucrose. When, this synthetic aC powder was mixed with sulfur and heated to 400 °C, the product obtained also exhibited traces of two SC phases at $T_c = 17$ K and at 42 K [12]. (iii) The third system we studied comprise thin composite films of granular aC and tungsten (aC-W) prepared by Electron-Beam Induced Deposition (EBID). After treatment with sulfur at 250 ºC the resultant aC-W-S films showed clear, yet still small, traces of SC at $T_c$ ~ 40 K [11].

The motivation for growing granular aC-W-S films was the hypothesis that tungsten inclusions, which are SC (with $T_c$ values that may vary significantly, see below), may increase the phase coherence between the localized SC islands in the aC matrix. This effect is anticipated to raise the $T_c$ and the SC volume fraction of sulfur treated film [11], akin to the $T_c$ enhancement effect found in bilayers of underdoped and overdoped $La_{2-x}Sr_xCuO_4$ high temperature cuprate superconductors [13]. The above three systems studied by us prove unambiguously the existence of SC in sulfur-doped aC materials. Our results are in good agreement with other experiments on related graphite-based systems, and corresponding theoretical works, implying that structural disorder, topological defects, sulfur atom adsorption or phosphorous ion implantation, trigger superconducting instabilities in graphite [14-16]. In particular, the experiments corroborate theoretical predictions, suggesting that both adsorbed sulfur and structural disorder can locally induce extra carriers (doping effect) into graphite/graphene and therefore trigger or enhance SC [17]. It should be pointed out, however, that in all these experiments [1,3,11-12,14] the SC phase was localized (not percolating) and the SF obtained were very low and the $T_c$ values largely vary between the different samples.

As mentioned above, only *traces* of SC around 40 K were observed in the EBID grown aC-W-S thin film. This brief report is focused on further experimental evidence which aim to prove (i) the reproducibility of the EBID process and (ii) that the SF obtained is very sensitive to the growing procedure of the film. Importantly, the magnetization curves presented here show a sharper SC transition compared to the data presented in Ref. 11, and, correspondingly, the estimated SF is enhanced by an order of magnitude.

**2 Experimental details**

The granular aC-W thin films were grown by employing the EBID technique as described in details in Refs. 11 and 17. EBID is a high-resolution single step fabrication process where an electron beam dissociates organo-metallic precursor molecules adsorbed to the surface of a sample, resulting in a granular metallic film. Here we used a $W(CO)_6$ precursor to deposit the blank granular aC-W film (labeled as 'blank') on the surface of a thermal oxide layer (25 nm thick) grown on a p-type Si wafer. To this end, the $W(CO)_6$ precursor was introduced into the chamber of a scanning electron microscope (e-line, Raith), but the parameters used (electron-beam energy, working distance, pressure) were varied compared to those described in Refs. 11 and 17, in order to optimize the SC properties of the resultant samples. Then, the sulfur composite (aC-W-S) was prepared by heating the blank film with sulfur powder (Aldrich Chemical Company, Inc.) in evacuated quartz tube at 250 °C for 24 h before cooling down to room temperature. Zero-field-cooled (ZFC) and field-cooled (FC) magnetization measurements at low applied magnetic fields ($H$) in the temperature range 5K<$T$<150 K have been performed using a commercial (Quantum Design) superconducting quantum interference device (SQUID) magnetometer, with samples mounted in gel-caps. Prior to recording the ZFC curves, the SQUID magnetometer was always adjusted to be in a "*real*" $H = 0$ state.

## 3. Results and discussion.

Figure 1 presents temperature magnetization curves measured at 89 Oe on an aC-W-S film and, for comparison, on the corresponding blank (aC-W) film before the sulfur treatment. The blank film shows positive temperature-independent magnetization, exhibiting no sign of SC in the measured temperature range. The plotted ZFC and FC branches for the aC-W-S film were obtained after subtracting the blank magnetization from the measured data. The onset of SC at $T_c$ =34.4 K is readily observed. Below $T_c$, the ZFC magnetization is negative (diamagnetic), as expected for a SC material, where the diamagnetism originates from screening super-currents. The clear negative FC magnetization, which is associated with the magnetic flux expulsion due to the Meissner effect (ME), serves unambiguously as the finger-print for a superconducting phase.

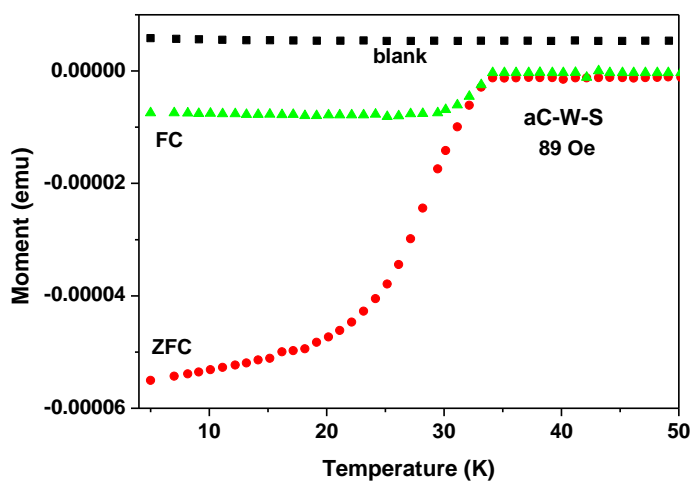

**Fig. 1** ZFC and FC temperature-dependent magnetization curves measured at 89 Oe on an aC-W-S film, plotted along with the magnetization of the corresponding blank (aC-W) film.

As expected, for higher applied fields (up to $H$ = 2 kOe) $T_c$ shifts to lower temperatures with a rate of $dH/dt$ = 0.45 kOe/K (not shown here). Due to the field dependence of the positive blank magnetization, determination of the onset ($H_{C2}$) at higher applied fields was not accurate. Since neither the weight nor the composition of the SC phase is known, any estimation of the SF from the ZFC branch is misleading. Nevertheless, the smallness of both SF and ME values, as well as the smeared ZFC signal, all suggest an inhomogeneous superconducting state in the aC-W-S film. Since the blank and aC-W-S films differ only in their sulfur contents, it is reasonable to assume that the observed SC state for aC-W-S (Fig.1) is triggered by the sulfur dopants. Due to its non-percolative small SC fraction, neither resistivity nor STM measurements have been performed. However, the SF presented here is an order of magnitude higher than that reported in Ref. 11.

Two further indications for the existence of SC in the aC-W-S films are exhibited in Figs 2-3. (i) The remnant magnetization that becomes zero very close to $T_c$, as demonstrated by Fig. 2. The remnant was recorded after cooling the aC-W-S film under 5 kOe from 38 K, when the applied field was switched off at 5 K. It is reasonable to relate this remnant to the trapped magnetic flux (vortices) in the material.

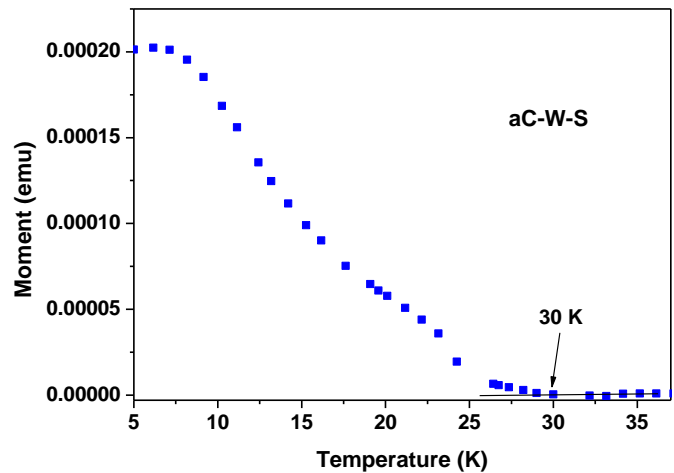

**Fig. 2** The remnant magnetization of aC-W-S measured after cooling from 38 K under an applied field of 5 kOe that was turned off at $T$ = 5 K

(ii) The two typical hysteresis loops measured at 5 and 10 K (Fig. 3). At low fields the *dM/dH* plots are linear. The deviations from linearity which reflects the lower critical field ($H_{c1}$) are: ~ 600 and ~400 Oe for $T$ = 5 and 10 K, respectively.

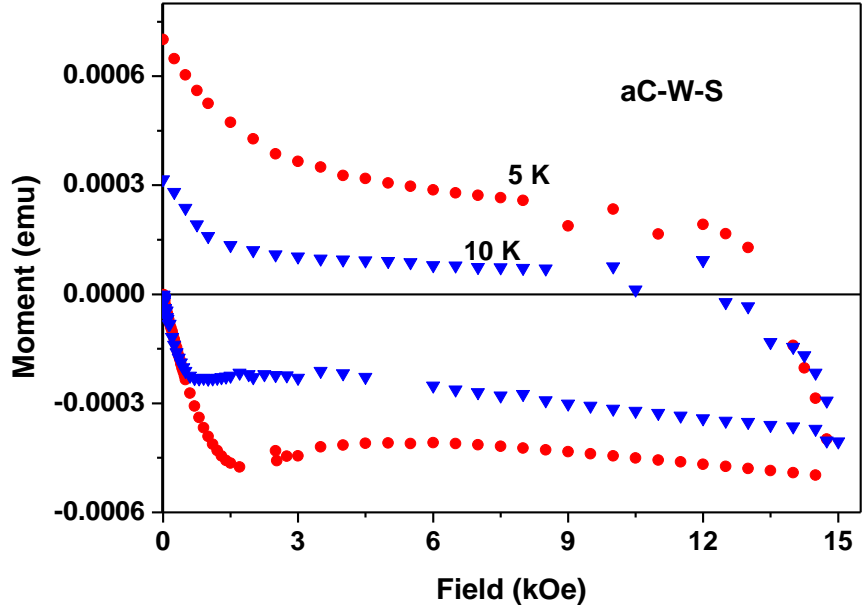

**Fig. 3** Hysteresis loops of the aC-W-S film measured at 5 and 10 K.

Figures 1-3, prove definitely the existence of a localized SC phase with $T_c$ = 34.4 K in the aC-W-S film. Although $T_c$ found here is smaller than that found in our previous EBID prepared samples [11], here the SC features in the magnetization curves are significantly clearer and more conclusive. In particular, the SC transition is sharper, and even more significantly, the ME is clearly observed in the FC branch. This may possibly stem from the larger concentration of W in the aC-W-S matrix prepared with the more optimized deposition conditions used here. The W islands probably do not contribute by themselves to the onset of SC in our films, since it appeared only after the sulfur treatment. In addition, the maximal reports critical temperature of any composite comprising W and a non-SC material did not exceed 6.5 K. The $T_c$ of bulk α-phase W is ~15 mK, whereas in the less stable β-phase the reported $T_c$ values are in the range 1-4 K [18]. Significant enhancement of $T_c$, up to ~ 6.5, was found in films deposited using focused ion-beam (FIB), also using W(CO)$_6$ [18], whereas in amorphous W-Si multilayers the maximal achieved $T_c$ is 4.2 K [19].

These transition temperatures are significantly smaller than those found here (Fig. 1) and in Ref. 11. It thus appears that the observed superconductivity is very likely due to a small fraction of an aC-S phase developed after the sulfur treatment, whose exact composition is not yet known. This conclusion is corroborated by the observation that also in the two other systems we have studied ('old' commercial and synthesized aC powders) SC emerged only after mixing and heating with sulfur. The role of the W inclusions (which do have a pairing potential) in the films studied here is probably in enhancing the phase coherences by Josephson coupling nearby localized SC regions. As is well known, superconductivity involves two imperative: the binding of electrons into pairs, which in our case may take place in the aC-S regions, and the establishment of phase coherence between them, presumably mediated more effectively by the W inclusions. Further optimizing the 'blank' aC-W film growth procedure, possible by using FIB (for which global $T_c$ was achieved at ~ 6.5 K) instead of EBID, may yield high-temperature percolative SC after sulfur treatment.

Our investigations trigger further questions related to the exact mechanism leading to SC in aC-S, and the means to enhance $T_c$ and the volume fraction to the extent of achieving macroscopic-percolative high-temperature SC in such materials. These are still the early days of the high-$T_c$ SC in carbon-based materials, and we hope that the present work, although showing only signatures of localized SC, as well as the exciting observation of localized SC at room temperature in water-treated graphite [9] would stimulate future studies in this direction. We note that SC in many carbon-based materials systems is a meta-stable phenomenon, therefore a lengthy, systematic experimental work, aiming to increase the reproducibility as well as the SC volume fraction in doped aC, is urgently needed.

**Acknowledgments**: The research is supported in parts by the Israel Science Foundation (ISF, Bikura 459/09), the German-Israel DIP program and by the Klachky Foundation for Superconductivity. O.M. thanks support from the Harry de Jur Chair of Applied Science.